# Cause of East-West Earth Asymmetry

by
**Giancarlo Scalera**

*INGV – Istituto Nazionale di Geofisica e Vulcanologia (retired)*
*Via di Vigna Murata 605 – 00143 Roma, Italy – giancarlo.scalera@ingv.it*

1. **Introduction**

Many efforts to explain some asymmetric characteristics of our globe and of the global tectonics have been made (Bostrom, 1971; Stevenson & Turner, 1977; Marotta & Mongelli, 1998; Doglioni et al., 1999; Carminati & Doglioni, 2012) most of them using the implicit or explicit assumptions of plate tectonics geodynamics.

A non exhaustive list of asymmetries of the Earth is: The magnetic polarity (William Gilbert, 1600); The land-hemisphere and the water-hemisphere; Southern tips of the continents (Francis Bacon, 1620, Mantovani, 1909; Barnett, 1969; Carey, 1996; among others); Larger extension of expanding mid-oceanic ridges on the southern hemisphere. South-eastward trend of younger ages in the long Pacific seafloor volcanic chains; A larger width of the seafloor isochrones bands on the Nazca region; A pear-shaped Earth; etc. Many other additional asimmetries have been described (see for a review: Carey, 1959, 1963) .

It is a few decades that the different slopes of the Wadati-Benioff zones oriented towards the east and west has been enclosed in the list of asymmetries of the Earth. Namely, the alleged subductions under the Americas have an angle of about 30°, while under the east Pacific coasts (Asia, Japan) the angle is steeper (Luyendyk, 1970; Isacks & Barazangi, 1977; Riguzzi et al., 2010; and many others). The cause of this difference has been identified in the tidal drag that would cause a global shift of the lithosphere towards west (the so called "westward drift"), but many objections can be advanced and other explanations are possible. A more complete and realistic view about the global tectonics asymmetries can be built only on the basis of the general geodynamics of the expanding Earth.

2. **The Coriolis inertial effect**

The unavoidable existence of the Coriolis fictitious force on a rotating Earth (Gerkema & Gostiaux, 2012; Gerkema et al., 2008) has inspired several authors to search for the possible effects of this sollecitation on the surface observable tectonic features. Van Bemmelen (1966, 1971) considered the Coriolis effect having an important role in his megaundation conception; Rance (1967) has searched for global lineaments of torsional origin on the physiographic map of the Pacific Ocean; possible observable effects on the surface of the Coriolis force on the tangential mantle flows has been described by Howell (1970) as clockwise and counterclockwise rotation of the seafloor transform faults on northern and southern hemisphere respectively; Kane (1972), has proposed a differential effect of the continents rotational inertia as a cause of plate tectonics; Hughes (1973) studied the possible effects on mantle convection of the Coriolis effect; Storetvedt (1992) used paleomagnetic data to build a view in which rotation of the plates are consequence of Earth's rotational effects; Pan (1993) repropose the possibility of influences of internal torques on tectonic features and polar motion, and Donescu & Munteanu (2011) confirm his argument.

In the opposite party was Jeffreys (1928) and many others up to our days (Jordan, 1974; Ranalli, 2000; Ricard, 2007; Doglioni et al., 2011), whose main argument is the extremely large viscosities of mantle, and the consequent assumption that the inertial terms in the Navier-Stokes equations are negligible. In a Coriolis-effect-free Earth's mantle, the westward drift of tidal origin of the lithosphere has been adopted by main stream (Bostrom, 1971; Doglioni et al., 2011; among many others) as explanation of the East-West asymmetry of the Wadati-Benioff zones, with additional assumptions. But the same argument of high viscosity can be used to reject the westward drift because the negligible value of tidal force in comparison to viscous friction (Jordan, 1974; Ranalli, 2000; Caputo & Caputo, 2012).

In the first years of plate tectonics the hypothesis was proposed that the plates could be decoupled from underlying mantle at level of asthenosphere, but Jordan (1974) proved that the depths of oceanic and continental lithospheres are very different and that the roots of continents can be detected up to 350 km. The consequent undulations of the ideal surface that defines the roots of oceans and continents do not allow for a tidal westward drift on it. Plate tectonicists have resolved the problem by hypothesizing a thin low viscosity layer at the depth of about 400 km, immediately upon the transition zone but still not observed by seismology (Caputo & Caputo, 2012). Besides the lacking of evidence in favor of this thin layer, evidence exist of regional upwelling of the 400 km discontinuity. In the Mediterranean region, a rising of the discontinuity was detected by higher modes of surface waves along an Istria Peninsula (north Adriatic sea) to Sardinia path (Scalera et al., 1981a, 1981b) and confirmed by seismic tomography (Piromallo & Morelli, 2003). The evidence of strong undulations of transition zone produces consequent problems in the hypothesized thin low viscosity layer, which could be uplifted by the upwelling transition zone or cutted and interrupted by it. Again a great difficulty results on the horizontal motion of the plates needed in the explanation of the East-West asymmetry without Coriolis effect.

Although the mutual importance of all the forces acting on the mantle materials must first be assessed, it is important to note that the observable facts seem to indicate a non-negligible action of the inertial Coriolis effect triggered by the Earth's rotation. Indeed, the Earth is rotating from West toward East and consequently each vertical motion directed from the depths towards the surface should be deviated away from the perfect verticality by a sufficiently strong Coriolis force, undergoing a bending toward West (Fig. 2b).

Obviously, the extrusion of mantle materials does not occur along perfectly vertical tracks, but following already existing discontinuity lines. For example the emerging flows adjacent to the western continental margins must have born already with a bending to west, and a more pronounced bending will be the result of the long time of action of the inertial force. If, on the contrary, the flows are near the eastern continental margins, starting already with an eastward bending, the Coriolis force will make them more vertical. The Pacific ocean-floor volcanism is more developed on the western side of the median ridge, and also this can be argued as caused by the prolonged westward action of the Coriolis force that possibly is able to detach "macro-drops" of rising materials and to lead them along more west directed bending paths.

Also the asymmetric topography across the rift zones, the compositional, thermal and density asymmetries (Panza et al., 2009, 2010; Doglioni et al., 2011), could find an integrated explanation in which the first cause is the Earth's rotation and the consequent inertial forces. In the same way that the gravity force operates as a sort of filter that drives the lighter compounds towards the surface and the heavier ones towards the geocenter, the Coriolis force could constitute an "East-West filter". It could drive the heavier minerals towards West, where they appear as constituting a "*fertile mantle*", while a "*depleted mantle*" is the result to East.

In the following of the present paper I will discuss the conditions allowing or forbidding the Coriolis effect, but before to deal with the Coriolis effect, a reflection has to be made about the possibility of motion of the mantle as a fluid, namely of convective motion in the mantle.

*The Reynold Number* – In fluid mechanics, the Reynolds number $N_{Rey}$ is a dimensionless number that gives a measure of the ratio of inertial forces to viscous forces and consequently quantifies the relative importance of these two types of forces for given flow conditions:

$$N_{Rey} = \frac{V \cdot L}{\nu}$$

where $V$ is the mean velocity of the fluid, $L$ is the characteristic length of the geometry (motions of mantle materials on lenght of undreds of kilometers), and $\nu$ is the kinematic viscosity. The higher the Reynolds number is, the more turbulent the flow will be: If $N_{Rey}$ < 2000 the flow is laminar; If 2000 < $N_{Rey}$ < 4000 it is called transition flow; If $N_{Rey}$ > 4000 the flow is turbulent.

Recalling that kinematic viscosity $\nu$ is the ratio of dynamic viscosity to density $\nu = \mu/\rho$, we take the following values: $V = 1$ cm/y $\approx 3.17 \cdot 10^{-10}$ m/s ; $L = 10^5$ m ; $\mu_{UM} \approx 10 \cdot 10^{19}$ Pa s, = kg/(s·m) (Harig et al., 2010); $\rho = 3.3$ g/cm$^3$=3.3·10$^3$ kg/m$^3$. Consequently the Reynolds number for the Earth's upper mantle is $N_{Rey} = 2.1 \cdot 10^{-21}$ which is a very little value indicating a slow laminar flow. However, all the researches on the mantle convection assume as starting point a layered non-expanding Earth, which may be a model far from reality.

*The Rossby Number* – It is also important to know if the role of Coriolis effect is important with respect to other inertial forces. It is sufficient to evaluate the Rossby number:

$$N_{Rossby} = \frac{V}{L \cdot f}$$

with $V$= typical velocity of the involved material; $L$= typical length on which the phenomenon develops; $f = 2\omega \cdot \sin\phi$ = Coriolis parameter ($\phi$ = latitude). The value of $N_{Rossby}$ must be very littler than 1.0 to assure that Coriolis effect is important. In the case of motions of mantle materials on length of hundreds of kilometers we can assume $\omega \approx 10^{-5}$ rad/s, $L = 10^5$ m, $V = 1$ cm/y $\approx 3.17 \cdot 10^{-10}$ m/s. With the same values for $L$ and $V$ adopted in the preceding Reynold number and $\omega = 10^{-5}$ rad/s it results:

$$N_{Rossby} \approx \frac{3.2 \cdot 10^{-10}}{\sin\phi}$$

a value ever extremely little, except very near to the equator. Because we are not treating aeronomical problems (which have material motions tangential to the Earth's surface) but we are mainly interested to radial displacements, the $\phi$ must be taken as the colatitude, and the value overcome the unity only near the poles. Then the Coriolis effects are dominant on other inertial forces, but our judgment should not be hasty about their real importance, because the existence of strong viscous friction can mitigate or make them negligible.

*Ekman number* – In a fluid, the Ekman number is the ratio of the viscous forces to the Coriolis fictitious forces. It has different definitions but the classic one is

$$E_K = \frac{v}{2L\omega \sin \phi}$$

Assuming for $L$ and $\omega$ the same values as in the above discussed Rossby number, and for ν, the kinematic viscosity, the same value as in the preceding Reynold number, the resulting value is $E_k \approx 10^9$ which mean an inescapable prevalence of the viscous forces on the Coriolis fictitious forces. The trajectories that Coriolis force would impose (Fig. 2a) in a non-viscous fluid (Paldor & Killworth, 1988) cannot be followed because the viscous friction. Then, in the mantle, at least for motions tangential to the sphere, the effects of the Earth's rotation can be neglected.

## 3. Round-trip or one-way tickets

There are at least three main version of the expanding Earth concept: i) The first version accepts the hypothesis of subduction and possibly of the convective flows (Owen, 1976, 1983, 1992; Perin, 2012). It is only a question of a non-equilibrium between the amount of subducted materials and new materials upwelled at the mid oceanic ridges – the last ones are hypothesized to prevail. ii) The more radical second version does not admit the existence of the subduction (Carey, 1975a, 1975b, 1976, 1996; Vogel, 1984; Maxlow, 2005). iii) A third version does not admit the existence of the large scale subduction, but a limited amount of regional underthrusts and overthrusts (few tens of km; Scalera 2007a, 2010, 2012) is admitted, in agreement with geological observations.

In plate tectonics the kinematics of the plates has been completed by a geodynamics that attributes the cause of continental drift to the convective cycles of the mantle and to other forces such as slab pull and slab-push. Instead, in the expansion global tectonics the main flows of the mantle materials are not necessarily moving along closed cycles of convection cells (Fig. 2a), but can be mainly extrusion flows along surfacewards paths. These non-cyclic surfaceward directed flows (one-way tickets instead of round-trips) must undergo the laws of the classical physics of fluid-dynamics. Being the Earth a rotating body, the inertial forces, like the Coriolis ones, must be present and, if sufficiently strong, should be considered among the factors influencing the final pattern of the flows.

In an expanding Earth, at least in the upper mantle, the radial flows of mantle materials are not necessarily slowed by viscous resistances. As explained in other papers (Scalera, 2003, 2010, 2012) the expansion can favor the isostatic rising of very deep material along huge and deep geofractures, which morphology – revealed by catalogues of relocated hypocenters (Engdahl et al., 1998) – resembles trees or smoke plumes enlarging and assuming the shape of great *calderas* (like the South Tyrrhenian one) towards the surface. Sudden motion, in the upper mantle, is revealed by earthquakes.

The isostatic rising of these materials can nullify the rising of deep materials due to thermal convection, in the sense that the progressive enlarging size – triggered and driven by global expansion – of the 'room' in which the rising materials are moving may not allow the onset of the convective circulation. In this room the velocity of rising is not constant but irregular and mainly impulsive, the rising episodes coinciding with changing of phase, and its range can be reasonably assumed as equal to the sliding velocity of the two sides of a fault during an earthquake, $V = 1$ m/s – 10 m/s.

With $\omega = 7.27 \cdot 10^{-5}$ rad/s, in the case of $V = 1.0 – 10.0$ cm/y = $3.17 \cdot 10^{-9} – 3.17 \cdot 10^{-10}$ m/s, the value of the Coriolis force $F_{Cor} = 2\rho\omega V$ Nw/m$^3$ (Ricard, 2007) is:

$$F_{Cor} = 2\rho\omega V \text{ Nw/m}^3 = 2 \cdot 3.3 \cdot 7.27 \cdot 10^{-5} \cdot V = 0.15 \cdot 10^{-9} – 1.5 \cdot 10^{-9} \text{ Nw/m}^3,$$

But I can compute that in the case of impulsive velocities of V = 1 m/s –10 m/s

$$F_{Cor} = 2\rho\omega V \text{ Nw/m}^3 = 2 \cdot 3.3 \cdot 7.27 \cdot 10^{-5} \cdot V = 0.48 – 4.8 \text{ Nw/m}^3,$$

which is more than 10 order of magnitude greater than in the case of the convective slow laminar flow. Then, it cannot be excluded the possibility of a deflection of the vertical sudden flows.

A comparison with the centrifugal force $F_{Cen}$ is also useful:

$$F_{Cen} = \omega^2 \cdot L \text{ Nw/m}^3 \approx 5.3 \cdot 10^{-4} \text{ Nw/m}^3,$$

a value some order of magnitude less than the Coriolis force in this impulsive case. The centrifugal force is little but is able to deform the Earth's shape to an oblate ellipsoid, and this is additionally in favor of the possibility for the Coriolis force to deform the path of the impulsive rising of mantle material. Obviously we cannot expect that the surfaceward motion ever occurs with an impulsive mechanism, but – in the impossibility to know the percent of the path performed in slow or impulsive way – a not negligible contribution of impulsive risings must be assumed.

## 4. Evidence

Evidence that the subductive dynamics on the Wadati-Benioff zone (WBZ) is invalid are coming from coseismic phenomena of the recent great and shallow earthquakes (Sumatran quake: Han et al., 2006; Scalera, 2007b) (Honshu quake: Han et al., 2011; among others).

The great Sumatra earthquake has caused a sudden displacement of the instantaneous rotation pole of the Earth (Bianco, 2005). Scalera (2007b, 2012) has evidenced that the rotation axis moved following the meridian of the epicenter, going nearly 3 milliarcsec (≈ 10.0 cm) farther from the epicenter (Fig. 3a). The data of the daily averages (Fig. 3a) show some indications for a mass displacement many hours before the quake. Rational mechanics rules (Schiaparelli, 1891) make clear that additional mass has been emplaced in the earthquake zone (Scalera, 2007b), following a mechanism of extrusion. The data of the GRACE satellites (Han et al., 2006) show variations of surface gravity of -15 μGal east of the Sunda trench, and a symmetrical anomaly of +15 μGal west of the trench. These anomalies does not fit a fault dislocation without a substantial lateral and vertical expansion of the oceanic crust.

The great earthquake of Honshu Tohoku (11 March 2011; Mw=9.1) has produced similar effects (Fig. 3b). Instead of a coseismic displacement of the instantaneous rotation pole of 14 cm toward 135°E as forecasted by Gross (see Buis comment, 2012) using the Dahlen (1971) dislocation model, a tendency of a little displacement toward an opposite direction (away from the Honshu hypocentral region) can be deduced (Scalera, 2012). Also in this case an extrusion of material is favored and the gravimetric data have confirmed (Han et al., 2011; Zhou et al., 2012). Similar GRACE results and interpretations have been published for the Maule quake (27 February 2010; Mw=8.8) (Heki & Matsuo, 2010).

Already since many years, geodetic GPS networks have given precise indications of what actually takes place on the active margins. The data collected to date for the active margins of Sumatra, Japan, South America and so on, reveal a coseismic deformation of distensional nature (Chlieh et al., 2007; Lee et al., 2008; among others).

### 5. Conclusion

The conclusion is that at least two estreme magnitude events have provided astrogeodetic evidence not in agreement with plate tectonics, but more in accord with an expanding and *emitting* Earth view.

A simplistic evaluation of the regime of the convective motion in the mantle and of the order of magnitude of the involved forces (viscous, buoyancy, inertial) hastily judges as negligible the role of the Coriolis effect in producing the observed slope differences of the Wadati-Benioff regions. Instead, it is possible to show that changing the assumptions implicit in the adopted geodynamics theory, or in other words, by adopting a different theory of global geodynamics, the role of the fictitious inertial forces can become substantial. In a different framework in which sudden extrusions of mantle materials occur by local phase change toward a more unpacked lattice, the value of the Coriolis fictitious force can rise of several magnitude orders, becoming the main cause of the east-west asymmetry of the Wadati-Benioff zones, which might be ascribed entirely to internal causes of the planet (its rotation and geodynamics) and not to external causes (influence of other celestial bodies).

**Aknowledgements**

This work was written and improved during the few months preceding and following my retirement from INGV, to which my thanks go for having allowed to develop my research for thirty years in a field strongly opposed with respect to the "mainstream".

FIGURE CAPTIONS

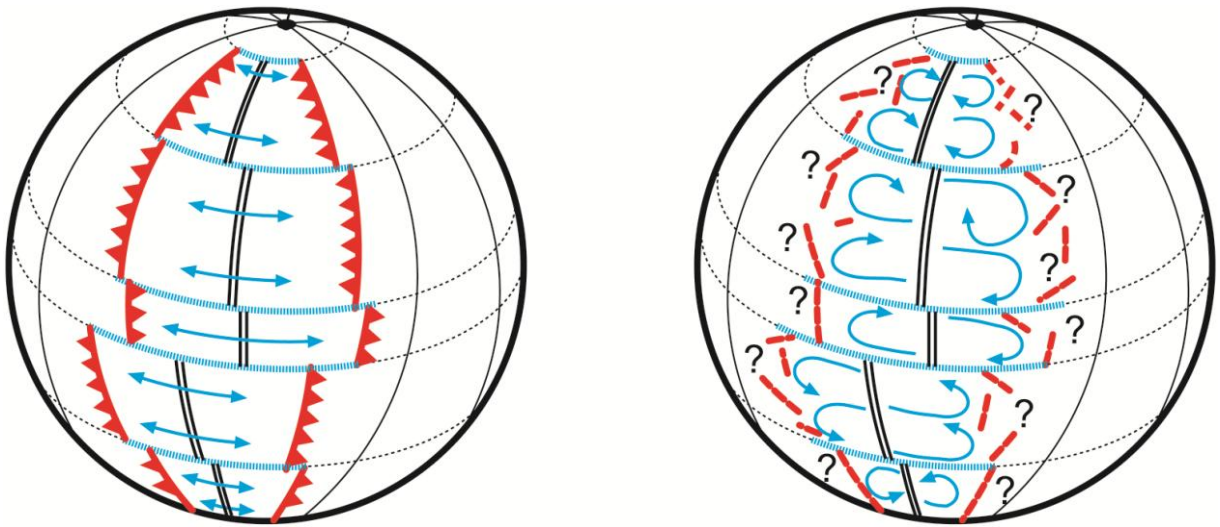

FIG. 1 – a) The plate tectonics representation of the plate motion does not allows Coriolis effects. Subduction zones are well defined at the leading edges of the plate motions. – b) If inertial Coriolis effects would be present the plate paths would be circular, with a problematic definition of the subductive margins, but no evidence of this pattern is observable. This pattern is not possible on an expanding Earth because of the extremely limited or completely absent horizontal motions.

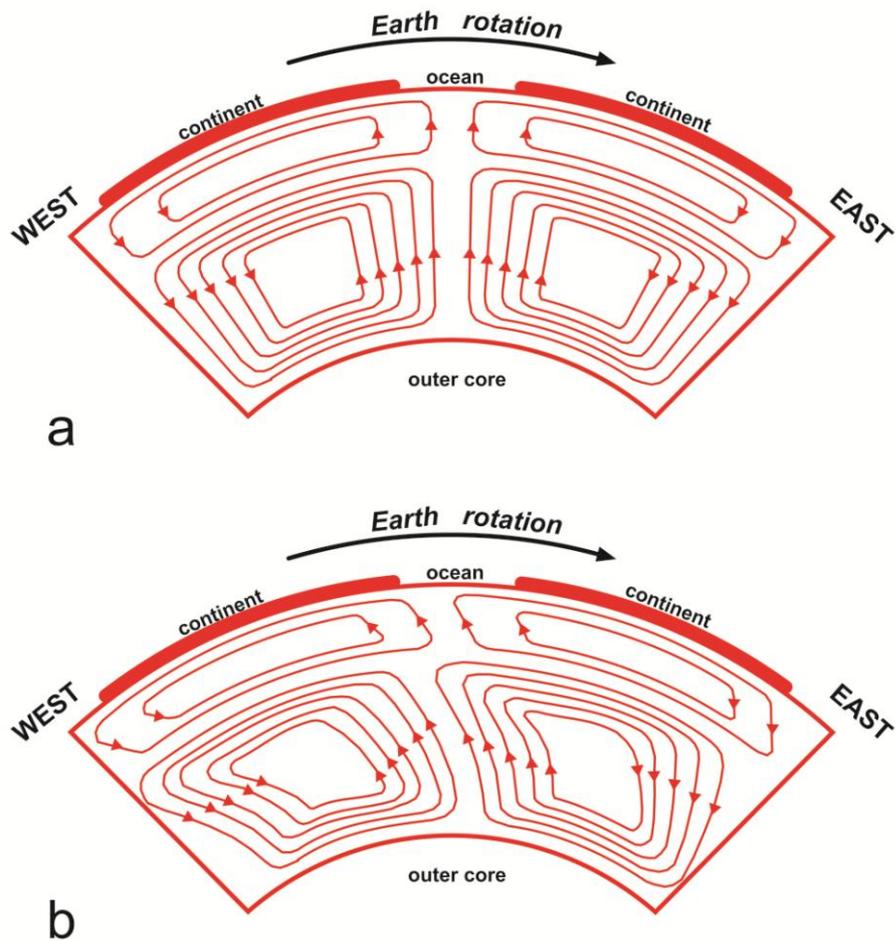

FIG. 2 – a) Convective motions in the Earth's mantle divided in upper-mantle and lower-mantle cells. In this representation Coriolis inertial effect is not taken into account because – following the main stream conception – only a laminar flow of few cm/yr are believed to occur. The Ekman ratio $E_k \approx 10^9$ does not allows significant Coriolis effects. – b) The convective motions of a rotating Earth could be deformed by the Coriolis effect in their upward and downward flows if impulsive motions occur. During earthquakes the mantle materials can slip with velocities in the range $V= 1.0$ - $10.0$ m/s. In this case inertial forces cannot be neglected in comparison to viscous friction. In the expanding Earth framework the surfacewards impulsive motions are very likely to occur.

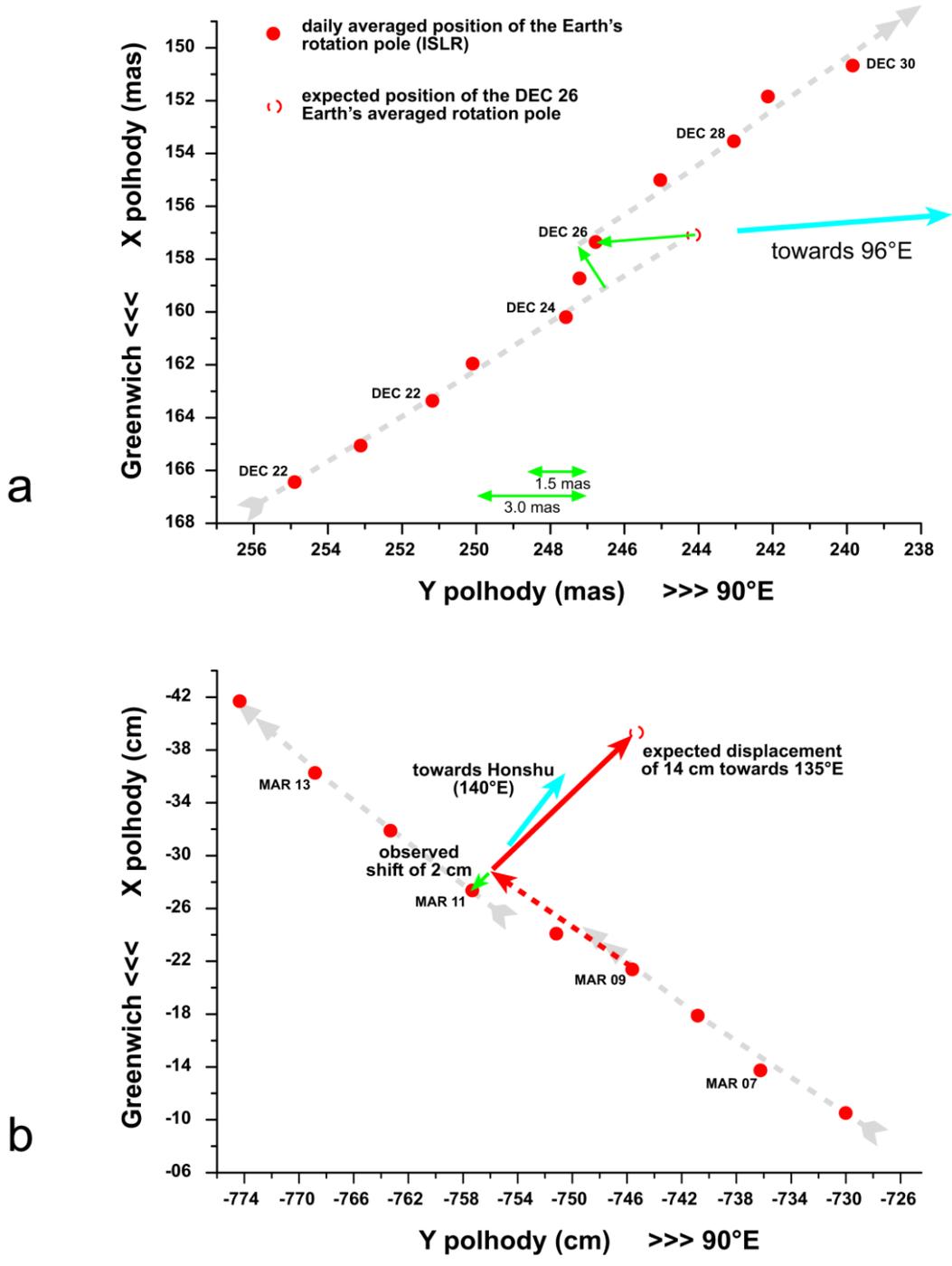

FIG. 3 – a) The shift of the Earth's rotation axis produced by the great Sumatran quake of December 26 2004. The polhody from December 20 to 30 is plotted (daily averages; data ISLR provided by IERS). The direction of the axis shift is toward an azimuth opposite to the hypocentral zone azimuth and at odds with respect to the plate tectonics forecasting. The data of the daily averages show some indications for a mass displacement many hours *before* the quake – b) The shift of the Earth's rotation axis produced by the Honshu quake of 11 March 2014 (Mw=9.1). The polhody from March 6 to March 14 is plotted (daily averages by IERS web-site facilities). Albeit less clearly than in the Sumatran event, the direction of the axis shift is toward an azimuth opposite to the hypocentral zone azimuth and at odds with respect to the 14 cm plate tectonics forecasting.